\documentclass[a4paper,12pt,leqno]{article}
\usepackage{theorem}
\newtheorem{thm}{THEOREM}
\newtheorem{lem}[thm]{LEMMA}
\newtheorem{cor}[thm]{COROLLARY}
\newtheorem{pro}[thm]{PROPOSITION}
\theoremheaderfont{\scshape}

\newcommand{\ket}[1]{|#1\rangle}

\title{{\Large {\bf LIMIT THEOREMS AND ABSORPTION PROBLEMS FOR QUANTUM RANDOM WALKS IN ONE DIMENSION}
}}
\author{{\small By NORIO KONNO } \\
{\small
{\it Yokohama National University}}
}

\date{\empty }
\begin{document}
\maketitle

\par\noindent
\begin{small}
{\bf Abstract}. In this paper we consider limit theorems, symmetry of distribution, and absorption problems for two types of one-dimensional quantum random walks determined by $2 \times 2$ unitary matrices using our PQRS method. The one type was introduced by Gudder in 1988, and the other type was studied intensively by Ambainis {\it et al.} in 2001. The difference between both types of quantum random walks is also clarified.

\footnote[0]{
{\it Key words and phrases.} 
Quantum random walk, the Hadamard walk, limit theorems, absorption problems. 
}

\end{small}

\setcounter{equation}{0}
\section{Introduction}
\newcommand{\U}{\bar{U}}

The classical random walk (CRW) in one dimension is the motion of a particle located on the set of integers. The particle moves at each step either one unit to the left with probability $p$ or one unit to the right with probability $q=1-p.$ The directions of different steps are independent of each other. This CRW is often called the Bernoulli random walk. In the present paper, we consider quantum variations of the Bernoulli random walk and refer to such processes as quantum random walks (QRWs) here. Sometimes the QRW is also called the quantum walk.

\par
Very recently, considerable work has been done on discrete-time and continuous-time QRWs by a number of groups in connection with quantum computing. Examples include Aharonov {\it et al.} (2001), Ambainis {\it et al.} (2001), Bach {\it et al.} (2002), Brun, Carteret and Ambainis (2002a, 2002b, 2002c), Childs, Farhi and Gutmann (2002), Childs {\it et al.} (2002), D\"ur {\it et al.} (2002), Kempe (2002), Kendon and Tregenna (2002a, 2002b), Konno (2002a, 2002b), Konno, Namiki and Soshi (2002), Konno, Namiki, Soshi and Sudbury (2003), Leroux (2002), Mackay {\it et al.} (2002), Moore and Russell (2001), Severini (2002a, 2002b), Travaglione and Milburn (2002), Yamasaki, Kobayashi and Imai (2002). For a more general setting including quantum cellular automata, see Meyer (1996). The present paper is concerned only with the discrete-time case.

\par
QRWs behave quite differently from CRWs. It is well known that the probability distribution of a CRW is given by the form of a binomial distribution. The variance of the CRW increases linearly with the number of time steps. By contrast, the probability distribution of a QRW has a complicated oscillatry form and the variance of the QRW increases quadratically with the number of time steps (see Ambainis {\it et al.} (2001), Konno (2002a), for examples). Such behavior of the QRW is due to the interference between the separate paths of the walk.

\par
The QRW was introduced by Gudder (see Section 7.4 in his book (1988)) to describe the motion of a quantum object in discrete space-time (very recently, I happened to know this fact from Severini (2002a)). QRWs have been re-discovered and studied intensively by Ambainis {\it et al.} (2001) in the context of quantum computation recently. So for simplicity we call the former QRW {\it the G-type QRW}, and the latter QRW {\it the A-type QRW} respectively in this paper. Both QRWs are essentially same, however they have a difference. One of the main purpose of this paper is to clarify the different behaviour between the two types QRWs, for examples, limit distribution, symmetry of distribution, absorption probability.

\par
Here we present not only our recent results on limit theorems, symmetry of distribution, and absorption probabilities for A-type QRWs, but also the related results for G-type QRWs, by using our PQRS method based on combinatorics which is different from Fourier analysis. This paper is organized as follows. Section 2 treats the definitions of both A-type and G-type QRWs and explains the PQRS method. In Section 3, we give a new type of limit theorems for both types of QRWs. Section 4 is devoted to absorption problems for QRWs.

\section{Definition and PQRS Method}

The time evolution of the one-dimensional QRW is given by the following unitary matrix:
\begin{eqnarray*}
U=
\left[
\begin{array}{cc}
a & b \\
c & d
\end{array}
\right]
\end{eqnarray*}
\par\noindent
where $a,b,c,d \in {\bf C}$ and ${\bf C}$ is the set of complex numbers. So we have 
\begin{eqnarray*}
&& |a|^2 + |b|^2 =|c|^2 + |d|^2 =1, \qquad a \overline{c} + b \overline{d}=0, \\
&& c= - \triangle \overline{b}, \qquad d= \triangle \overline{a}
\end{eqnarray*}
where $\overline{z}$ is a complex conjugate of $z \in {\bf C}$ and $\triangle = \det U = ad - bc.$ We should note that the unitarity of $U$ gives $|\triangle|=1.$
\par
For the A-type QRW, each coin performs the evolution
\begin{eqnarray*}
&& \ket{L} \qquad \to \qquad U \ket{L}=a \ket{L} + c \ket{R}, \\
&& \ket{R} \qquad \to \qquad U \ket{R}=b \ket{L} + d \ket{R}
\end{eqnarray*}
at each time step for which that coin is active, where R and L can be respectively thought of as the heads and tails states of the coin, or equivalently as an internal chirality state of the particle. The value of the coin controls the direction in which the particle moves. When the coin shows L, the particle moves one unit to the left, when it shows R, it moves one unit to the right. (The G-type QRW is also interpreted in a similar way. We will give a precise definition of both types of QRWs later.) In this meaning, the QRW can be considered as a quantum version of the CRW with an additional degree of freedom called the chirality which takes values left and right. As for ways to regain the CRW from the QRW, see Brun, Carteret and Ambainis (2002a, 2002b, 2002c).

The amplitude of the location of the particle is defined by a 2-vector $\in {\bf C}^2$ at each location at any time $n$. The probability the particle is at location $k$ is given by the square of the modulus of the vector at $k$. For the $j$-type QRW ($j=A,G$), let $\ket{\Psi_{j,k} (n)}$ denote the amplitude at time $n$ at location $k$ where
\begin{eqnarray*}
\ket{\Psi_{j,k} (n)} = \left[
\begin{array}{cc}
\psi_{j,k} ^L (n) \\
\psi_{j,k} ^R (n)
\end{array}
\right]
\end{eqnarray*}
with the chirality being left (upper component) or right (lower component). Then the dynamics for $\ket{\Psi_{A,k} (n)}$ for the A-type QRW is given by the following transformation:
\begin{eqnarray}
\ket{\Psi_{A,k} (n+1)} = P_A \ket{\Psi_{A,k+1} (n)} + Q_A \ket{\Psi_{A,k-1} (n)}
\end{eqnarray}
where
\begin{eqnarray*}
P_A= 
\left[
\begin{array}{cc}
a & b \\
0 & 0 
\end{array}
\right], 
\quad
Q_A=
\left[
\begin{array}{cc}
0 & 0 \\
c & d 
\end{array}
\right]
\end{eqnarray*}
It is noted that $U=P_A+Q_A.$ The unitarity of $U$ ensures that the amplitude always defines a probability distribution for the location.
\par
On the other hand, the G-type of QRW can be determined by 
\begin{eqnarray*}
\ket{\Psi_{G,k} (n+1)} = P_G \ket{\Psi_{G,k+1} (n)} + Q_G \ket{\Psi_{G,k-1} (n)}
\end{eqnarray*}
where
\begin{eqnarray*}
P_G= 
\left[
\begin{array}{cc}
a & 0 \\
c & 0 
\end{array}
\right], 
\quad
Q_G=
\left[
\begin{array}{cc}
0 & b \\
0 & d 
\end{array}
\right]
\end{eqnarray*}
In this G-type case also, we see that $U=P_G+Q_G$, and the unitarity of $U$ ensures that the amplitude always defines a probability distribution for the location. In our one-dimensional setting, Gudder considered the following simple model (see Eqs. (7.33) and (7.34) in page 279 of Gudder (1988)) and computed $\ket{\Psi_{G,k} (n)}$ (in our notation) for this model (see Corollary 7.24 in page 285 of his book) by using a Fourier analysis which is different from our PQRS method:
\begin{eqnarray*}
U=
\left[
\begin{array}{cc}
a & ib \\
ib & a
\end{array}
\right]
\end{eqnarray*}
where $0<a<1$ and $b=\sqrt{1-a^2}$.

\par
The simplest and well-studied example of an A-type QRW is the Hadamard walk whose unitary matrix $U$ is defined by 
\begin{eqnarray*}
H = 
{1 \over \sqrt{2}}
\left[
\begin{array}{cc}
1 & 1 \\
1 & -1 
\end{array}
\right] 
\end{eqnarray*}
The dynamics of this walk corresponds to that of the symmetric CRW. In general, the following unitary matrices can also lead to symmetric QRWs:
\begin{eqnarray*}
U_{\eta, \phi, \psi} = {e^{i \eta} \over \sqrt{2}}
\left[
\begin{array}{cc}
e^{i(\phi + \psi)} & e^{-i(\phi - \psi)}  \\
e^{i(\phi - \psi)}  & -e^{-i(\phi + \psi)}  
\end{array}
\right] 
\end{eqnarray*}
where $\eta, \phi ,$ and $\psi$ are real numbers (see pp.175-176 in Nielsen and Chuang (2000), for example). In particular, we see $U_{0,0,0}=H$.  
\par
However symmetry of the Hadamard walk depends heavily on the initial qubit state, see Konno, Namiki and Soshi (2002). Another generalization of the Hadamard walk is:
\begin{eqnarray*}
H (\rho) = 
\left[
\begin{array}{cc}
\sqrt{\rho} & \sqrt{1-\rho} \\
\sqrt{1-\rho} & - \sqrt{\rho} 
\end{array}
\right] 
\end{eqnarray*}
where $0 \le \rho \le 1$. Note that  $\rho = 1/2$ is the Hadamard walk, that is, $H=H(1/2)$.

In the present paper, the study on the dependence of some important properties and quantities (e.g., symmetry of distribution, limit distribution, absorption probability) on initial qubit state is one of the essential parts, so we define the collection of initial qubit states as follows:
\[
\Phi = \left\{ \varphi =
\left[
\begin{array}{cc}
\alpha \\
\beta   
\end{array}
\right]
\in 
{\bf C}^2
:
|\alpha|^2 + |\beta|^2 =1
\right\}
\]

Let $X_{j,n} ^{\varphi}$ be the $j$-type QRW at time $n$ starting from initial qubit state $\varphi \in \Phi$ with $X_{j,0} ^{\varphi}=0$ for $j=A,G$. In our treatment of QRWs, as well as the matrices $P_j$ and $Q_j$, it is convenient to introduce
\[
R_A=
\left[
\begin{array}{cc}
c & d \\
0 & 0 
\end{array}
\right], 
\quad
S_A=
\left[
\begin{array}{cc}
0 & 0 \\
a & b 
\end{array}
\right]
\]
and 
\[
R_G=
\left[
\begin{array}{cc}
0 & a \\
0 & c 
\end{array}
\right], 
\quad
S_G=
\left[
\begin{array}{cc}
b & 0 \\
d & 0 
\end{array}
\right]
\]
We should remark that both A-type and G-type $P_j,Q_j,R_j, S_j \>(j=A,G)$ form an orthonormal basis of the Hilbert space $M_2 ({\bf C})$ which is the vector space of complex $2 \times 2$ matrices with respect to the trace inner product $\langle A|B \rangle = \>$ tr $(A^{\ast}B)$. Therefore we can express any $2 \times 2$ matrix $X$ conveniently in the form,
\begin{eqnarray} 
X = 
\hbox{tr} (P^{\ast} _j X)P_j + \hbox{tr} (Q^{\ast} _j X)Q_j + \hbox{tr} (R^{\ast} _j X)R_j + \hbox{tr} (S^{\ast} _j X)S_j 
\end{eqnarray} 
for each $j=A,G$. We call the analysis based on $P_j,Q_j,R_j,S_j \> (j=A,G)$ the PQRS method. 

The $n \times n$ unit and zero matrices are written $I_n$ and $O_n$ respectively. For instance, if $X = I_2$, then Eq. (2.2) gives
\begin{eqnarray} 
I_2 = \overline{a} P_j + \overline{d} Q_j + \overline{c} R_j + \overline{b} S_j
\end{eqnarray} 
for each $j=A,G$. The next table of products of $P_j,Q_j,R_j,S_j \> (j=A,G)$ is very useful in computing some quantities:
\par
\
\par
\begin{center}
\begin{tabular}{c|cccc}
  & $P_j$ & $Q_j$ & $R_j$ & $S_j$  \\ \hline
$P_j$ & $aP_j$ & $bR_j$ & $aR_j$ & $bP_j$  \\
$Q_j$ & $cS_j$ & $dQ_j$& $cQ_j$ & $dS_j$ \\
$R_j$ & $cP_j$ & $dR_j$& $cR_j$ & $dP_j$ \\
$S_j$ & $aS_j$ & $bQ_j$ & $aQ_j$ & $bS_j$ 
\end{tabular}
\end{center}
where $P_jQ_j=bR_j$, for example. We should remark that the algebraic structure for both types is the same.
\par
In order to consider absorption problems stated in Section 4, now we describe the evolution and measurement of the A-type QRW starting from location $k$ on $\{ 0, 1, \ldots , N \}$ with absorbing boundaries (for examples, see Ambainis {\it et al.} (2001), Bach {\it et al.} (2002), and Kempe (2002) for more detailed information). As for the G-type QRW, we can define its evolution and measurement in a similar fashion. 
\par
First we consider $N= \infty$ case. In this case, an absorbing boundary is present at location 0. The evolution mechanism is described as follows:
\par 
Step 1. Initialize the system $\varphi \in \Phi$ at location $k$.
\par
Step 2. (a) Apply Eq. (2.1) to one step time evolution. (b) Measure the system to see where it is at location 0 or not.
\par 
Step 3. If the result of measurement revealed that the system was at location 0, then terminate the process, otherwise repeat step 2.
\par
In this setting, let $\Xi^{(\infty)} _{A,k} (n)$ be the sum over possible paths for which the particle first hits 0 at time $n$ starting from $k$ for the A-type QRW. For example,
\begin{eqnarray*} 
\Xi^{(\infty)} _{A,1} (5) = P_A ^2 Q_A P_A Q_A + P_A ^3 Q_A ^2= (a b^2 c + a^2bd) R_A
\end{eqnarray*} 
The probability that the particle first hits 0 at time $n$ starting from $k$ is 
\begin{eqnarray*} 
P^{(\infty)} _{A,k} (n; \varphi) = | \Xi^{(\infty)} _{A,k} (n) \varphi |^2
\end{eqnarray*} 
So the probability that the particle first hits 0 starting from $k$ for the A-type QRW is 
\begin{eqnarray*} 
P^{(\infty)} _{A,k} (\varphi) = \sum_{n=0} ^{\infty} P^{(\infty)} _{A,k} (n; \varphi) 
\end{eqnarray*} 
\par
Next we consider $N < \infty$ case. This case is similar to the $N= \infty$ case, except that two absorbing boundaries are presented at locations 0 and $N$ as follows:
\par 
Step 1. Initialize the system $\varphi \in \Phi$ at location $k$.
\par
Step 2. (a) Apply Eq. (2.1) to one step time evolution. (b) Measure the system to see where it is at location 0 or not. (c) Measure the system to see where it is at location $N$ or not.
\par 
Step 3. If the result of either measurement revealed that the system was either at location 0 or location $N$, then terminate the process, otherwise repeat step 2.
\par
Let $\Xi^{(N)} _{A,k} (n)$ 
be the sum over possible paths for which the particle first hits 0 at time $n$ starting from $k$ before it arrives at $N$ for the A-type QRW. For example, 
\begin{eqnarray} 
\Xi^{(3)} _{A,1} (5) = P_A ^2 Q_A P_A Q_A =a b^2 c R_A
\end{eqnarray} 
In a similar way, we can define $P^{(N)} _{A,k} (n; \varphi)$ and $P^{(N)} _{A,k} (\varphi)$.

\section{Limit Theorem}
This section treats limit theorems for both A-type and G-type QRWs $X^{\varphi} _{j,n} (j=A,G).$ To study $P(X_{j,n} ^{\varphi} =k)$ for $n+k=$ even, it suffices to understand the following combinatorial expression. For fixed $l$ and $m$ with $l+m=n$ and $m-l=k$, we consider
\[
\Xi_j (l,m)= \sum_{l_i, m_i \ge 0: m_1+ \cdots +m_n=m, l_1+ \cdots +l_n=l} P_j ^{l_1}Q_j ^{m_1}P_j ^{l_2}Q_j ^{m_2} \cdots P_j ^{l_n}Q_j ^{m_n}
\]
since 
\[
P(X_{j,n} ^{\varphi} =k) = |\Xi_j (l,m) \varphi |^2
\]
Since $P_j, Q_j, R_j, S_j \> (j=A,G)$ are a basis of $M_2 ({\bf C})$, $\Xi_j (l,m)$ has the following form:
\[
\Xi_j (l,m) = p_{j} (l,m) P_j + q_{j} (l,m) Q_j + r_{j} (l,m) R _j+ s_{j} (l,m) S_j
\]
Next problem is to obtain explicit forms of $p_{j} (l,m), q_{j} (l,m), r_{j} (l,m),$ and $s_{j} (l,m)$. In the above example of $n=l+m=4$ case, we see that for $j=A,G,$ 
\begin{eqnarray*}
&& \Xi_j (4,0) = a^3 P_j, \quad
\Xi_j (3,1) = 2abc P_j + a^2b R_j + a^2c S_j, \quad \\
&& \Xi_j (2,2) =  bcd P_j + abc Q_j + abd R_j + bc^2 S_j, \\
&& \Xi_j (1,3) = 2bcd Q_j + bd^2 R_j + cd^2 S_j, \quad
\Xi_j (2,2) = d^3 Q_j
\end{eqnarray*}
So, for example,
\[
p_{j} (3,1)=2abc, \quad q_{j} (3,1)=0, \quad r_{j} (3,1)=a^2b, \quad s_{j} (3,1)=a^2c
\]
Note that $p_{j} (l,m), q_{j} (l,m), r_{j} (l,m), s_{j} (l,m)$ do not depend on the type $j$. 
\par
In $abcd=0$ case, the argument is much easier. So from now on we focus only on $abcd \not= 0$ case. In this case, the next key lemma is obtained by a combinatorial method. 
\par
\
\par\noindent
\begin{lem}
\label{lem:lem1} We consider both A-type and G-type QRWs with $abcd \not= 0$. Suppose that $l,m \ge 0$ with $l+m=n$, then we have 
\par\noindent
\hbox{(i)} for $l \wedge m (= \min \{l,m \}) \ge 1$,  
\begin{eqnarray*}
\Xi_j (l,m) 
= a^l d^m 
\sum_{\gamma =1} ^{l \wedge m} 
\left(-{|b|^2 \over |a|^2} \right)^{\gamma}
{l-1 \choose \gamma- 1} 
{m-1 \choose \gamma- 1} 
\times 
\Biggl[ {l- \gamma \over a \gamma } P_j + {m - \gamma \over d \gamma} Q_j + {1 \over c} R_j + {1 \over b} S_j 
\Biggr]
\end{eqnarray*}
\par\noindent
\hbox{(ii)} for $l (=n) \ge 1, m = 0$,  
\[
\Xi_j (l,0) = a^{l-1} P_j
\]
\par\noindent
\hbox{(iii)} for $l = 0, m (=n) \ge 1$,  
\[
\Xi_j (0,m) = d^{m-1} Q_j
\]
where $j=A,G$.
\end{lem}
The proofs of parts (ii) and (iii) are trivial. The proof of part (i) is based on a consequence of enumerating the paths of drift $l+m=n$. To do so, we consider the following 4 cases:
\begin{eqnarray*}
&&
\overbrace{P_jP_j \cdots P_j}^{w_1} \overbrace{Q_jQ_j \cdots Q_j}^{w_2} \overbrace{P_jP_j \cdots P_j}^{w_3} \cdots \overbrace{Q_jQ_j \cdots Q_j}^{w_{2 \gamma}} \overbrace{P_jP_j \cdots P_j}^{w_{2 \gamma+1}} \\
&&
\overbrace{Q_jQ_j \cdots Q_j}^{w_1} \overbrace{P_jP_j \cdots P_j}^{w_2} \overbrace{Q_jQ_j \cdots Q_j}^{w_3} \cdots \overbrace{P_jP_j \cdots P_j}^{w_{2 \gamma}} \overbrace{Q_jQ_j \cdots Q_j}^{w_{2 \gamma+1}} \\
&&
\overbrace{P_jP_j \cdots P_j}^{w_1} \overbrace{Q_jQ_j \cdots Q_j}^{w_2} \overbrace{P_jP_j \cdots P_j}^{w_3} \cdots \overbrace{Q_jQ_j \cdots Q_j}^{w_{2 \gamma}} \\
&&
\overbrace{Q_jQ_j \cdots Q_j}^{w_1} \overbrace{P_jP_j \cdots P_j}^{w_2} \overbrace{Q_jQ_j \cdots Q_j}^{w_3} \cdots \overbrace{P_jP_j \cdots P_j}^{w_{2 \gamma}}
\end{eqnarray*}
where $w_1, w_2, \ldots , w_{2 \gamma+1} \ge 1$ and $\gamma \ge 1$. The above each type of paths of $P_j$ and $Q_j$ corresponds to the each term of $P_j, Q_j, R_j, S_j$ respectively in the right hand side of the equation of part (i). As for the A-type case, the details of this proof appear in Konno (2002b). A similar proof also can be seen in Appendix A of Brun, Carteret and Ambainis (2002b). The proof of the G-type case is the same.

\par
By this lemma, the characteristic function of $X_{j,n} ^{\varphi} (j=A,G)$ for $abcd \not=0$ case is obtained. Moreover, the $m$th moment of $X_{j,n} ^{\varphi}$ can be derived from the characteristic function in the standard fashion. Here we give only the result of the $m$th moment.
\par
\
\par\noindent
\begin{pro}
\label{thm:pro2} 
We consider both A-type and G-type QRWs with $abcd \not= 0$.
\par\noindent
\hbox{(i)} When $m$ is odd, we have
\begin{eqnarray*}
E((X_{j,n} ^{\varphi}) ^m) 
&=& 
|a|^{2(n-1)}
\biggl[ - n^m \Gamma_j \\
&& + \sum_{k=1}^{\left[{n-1 \over 2}\right]}
\sum_{\gamma =1} ^{k} \sum_{\delta =1} ^{k}
\left(-{|b|^2 \over |a|^2} \right)^{\gamma + \delta} 
{k-1 \choose \gamma- 1} 
{k-1 \choose \delta- 1} 
{n-k-1 \choose \gamma- 1} 
{n-k-1 \choose \delta- 1} \\
&& 
\times {(n-2k)^{m+1} \over  \gamma \delta} 
\Delta_j
\Biggr]
\end{eqnarray*}
where 
\begin{eqnarray*}
&& 
\Gamma_A = \left(|a|^2 - |b|^2 \right)  
\left(|\alpha|^2 - |\beta|^2 \right) 
+ 2 (a \alpha \overline{b \beta} + \overline{a \alpha} b \beta ), 
\qquad
\Gamma_G = |\alpha|^2 - |\beta|^2, \\ 
&&
\Delta_A =
- \{ n( |a|^2 - |b|^2 ) + \gamma + \delta \} (|\alpha|^2 - |\beta|^2 ) 
+ \left( {\gamma +\delta \over |b|^2} - 2n \right) \Theta_A,
\\
&&
\Delta_G =
(\gamma + \delta - n) (|\alpha|^2 - |\beta|^2 ) 
- {\gamma +\delta \over |b|^2} \Theta_G,
\\
&& 
\Theta_A = a \alpha \overline{b \beta} + \overline{a \alpha} b \beta, 
\qquad
\Theta_G = a \beta \overline{c \alpha} + \overline{a \beta} c \alpha 
\end{eqnarray*}
\hbox{(ii)} When $m$ is even, we have
\begin{eqnarray*}
E((X_{j,n} ^{\varphi}) ^m) 
&=&  |a|^{2(n-1)} 
\Biggl[ 
n^m \\
&&
+
\sum_{k=1}^{\left[{n-1 \over 2}\right]}
\sum_{\gamma =1} ^{k} \sum_{\delta =1} ^{k}
\left(-{|b|^2 \over |a|^2} \right)^{\gamma + \delta} 
{k-1 \choose \gamma- 1} 
{k-1 \choose \delta- 1} 
{n-k-1 \choose \gamma- 1} 
{n-k-1 \choose \delta- 1} \\
&& 
\times {(n-2k)^{m} \over  \gamma \delta} 
\biggl\{
(n-k)^2 + k^2 - n (\gamma + \delta) + {2 \gamma \delta \over |b|^2} 
\biggr\}
\Biggr]
\end{eqnarray*}
\end{pro}
\par
\
\par\noindent
It should be noted that when $m$ is even, $E((X_{j,n} ^{\varphi})^m)$ is independent of initial qubit state $\varphi$ and types $j=A, G$. In particular, we use the above result of $m=1$ case in order to study symmetry of distributions for the QRWs. Moreover we have the following new type of limit theorem (as for the A-type QRW, see Konno (2002a)):
\par
\
\par\noindent
\begin{thm}
\label{thm:thm3} We consider both A-type and G-type QRWs with $abcd \not= 0$. Let $\Theta_A = a \alpha \overline{b \beta} + \overline{a \alpha} b \beta,$ and $
\Theta_G = a \beta \overline{c \alpha} + \overline{a \beta} c \alpha$. If $n \to \infty$, then 
\[
{X_{j,n} ^{\varphi} \over n} \quad \Rightarrow \quad Z_j ^{\varphi} \qquad (j=A,G)
\]
where $Z_j ^{\varphi}$ has a density 
\[
f_j (x; {}^t[\alpha, \beta])
= { \sqrt{1 - |a|^2} \over \pi (1 - x^2) \sqrt{|a|^2 - x^2}} 
 \left\{ 1- \left( |\alpha|^2 - |\beta|^2 + { \Theta_j \over |a|^2 } \right) x \right\} 
\]
for $x \in (- |a|, |a|)$ with 
\begin{eqnarray*}
&& E(Z_j ^{\varphi}) 
=
- \left( |\alpha|^2 - |\beta|^2 + { \Theta_j \over |a|^2 } \right) 
\times (1 - \sqrt{1 - |a|^2}) 
\\
&& E ((Z_j ^{\varphi})^2) = 1 - \sqrt{1 - |a|^2}
\end{eqnarray*}
and $ Y_n \Rightarrow Y$ means that $Y_n$ converges in distribution to a limit $Y$. 
\end{thm}
\par
\
\par\noindent
We remark that standard deviation of $Z_j ^{\varphi}$ is not independent of initial qubit state $\varphi ={}^t[\alpha, \beta]$. The above limit theorem suggests the following result on symmetry of distribution for both A-type and G-type QRWs. This is a generalization of the result give by Konno, Namiki and Soshi (2002) for the A-type Hadamard walk and by Konno (2002a) for the A-type QRW. For $j=A,G,$ define
\begin{eqnarray*}
\Phi_{j,s} &=&  \{ \varphi \in 
\Phi : \> 
P(X_{j,n} ^{\varphi}=k) = P(X_{j,n} ^{\varphi}=-k) \>\> 
\hbox{for any} \> n \in {\bf Z}_+ \> \hbox{and} \> k \in {\bf Z}
\} 
\\
\Phi_{j,0} &=& \left\{ \varphi \in 
\Phi : \> 
E(X_{j,n} ^{\varphi})=0 \>\> \hbox{for any} \> n \in {\bf Z}_+
\right\}
\\
\Phi_{j, \bot} &=& \left\{ \varphi = {}^t[\alpha, \beta] \in 
\Phi :
|\alpha|= |\beta|, \> \Theta_j =0 
\right\} \quad (j=A,G)
\end{eqnarray*}
and ${\bf Z}$ (resp. ${\bf Z}_+$) is the set of (resp. non-negative) integers.  For $\varphi \in \Phi_s$, the probability distribution of $X_{j,n} ^{\varphi}$ is symmetric for any $n \in {\bf Z}_+$. Using the explicit form of $E(X_{j,n} ^{\varphi})$ given by Proposition 2 (i) ($m=1$ case), we have 
\par
\
\par\noindent
\begin{thm}
\label{thm:thm4}  We consider both A-type and G-type QRWs with $abcd \not= 0$. Then for $j=A,G$ we have
\[
\Phi_{j,s} = \Phi_{j,0} = \Phi_{j, \bot}
\]
\end{thm}
From now on we give an outline of our proof of Theorem 3 (for more details, see Konno (2002b)). To do so, we introduce the Jacobi polynomial $P^{\nu, \mu} _n (x)$, where $P^{\nu, \mu} _n (x)$ is orthogonal on $[-1,1]$ with respect to $(1-x)^{\nu}(1+x)^{\mu}$ with $\nu, \mu > -1$. Then the following relation holds:
\[
P^{\nu, \mu} _n (x) = {\Gamma (n + \nu + 1) \over \Gamma (n+1) \Gamma (\nu +1)}
{}_2F_1(- n, n + \nu + \mu +1; \nu +1 ;(1-x)/2)
\]
where ${}_2F_1(a, b; c ;z)$ is the hypergeometric series and $\Gamma (z)$ is the gamma function. In general, as for orthogonal polynomials, see Koekoek and Swarttouw (1996). Remark that
\begin{eqnarray*}
&&
\sum_{\gamma =1} ^{k}
\left(-{|b|^2 \over |a|^2} \right)^{\gamma -1}
{1 \over \gamma} 
{k-1 \choose \gamma- 1}  
{n-k-1 \choose \gamma- 1} 
=
{1 \over k} |a|^{-2(k-1)} P^{1,n-2k} _{k-1}(2|a|^2-1)
\\
&&
\sum_{\gamma =1} ^{k}
\left(-{|b|^2 \over |a|^2} \right)^{\gamma -1}
{k-1 \choose \gamma- 1}  
{n-k-1 \choose \gamma- 1} 
= |a|^{-2(k-1)} P^{0,n-2k} _{k-1}(2|a|^2-1)
\end{eqnarray*}
By using the above equations and an expression of the characteristic function for $X^{\varphi}_{j,n}$, we obtain the next asymptotics of characteristic function $E(e^{i \xi X^{\varphi}_{j,n}/n})$: if $n \to \infty$ with $k/n=x \in (-(1-|a|)/2, (1+|a|)/2)$, then
\begin{eqnarray*}
&& E(e^{i \xi X_{j,n} ^{\varphi}/n}) 
\sim 
\sum_{k=1}^{\left[{n-1 \over 2}\right]} |a|^{2n - 4k -2}|b|^4 \\
&& \times 
\biggl[ 
\left\{ {2x^2-2x+1 \over x^2} (P^{1,n-2k} _{k-1})^2 - {2 \over x} P^{1,n-2k} _{k-1} P^{0,n-2k} _{k-1}+
{2 \over |b|^2} (P^{0,n-2k} _{k-1})^2 \right\} \cos ((1-2x) \xi) \\
&& 
- {(1-2x) \Gamma_j \over x^2 } 
(P^{1,n-2k} _{k-1})^2 
- 2 I_j \left( |\alpha|^2 - |\beta|^2 - 
{ \Theta_j \over |b|^2 } 
\right) P^{0,n-2k} _{k-1} P^{1,n-2k} _{k-1} \biggl\} 
i \sin ((1-2x) \xi)
\biggr]
\end{eqnarray*}
where $f(n) \sim g(n) $ means $f(n)/g(n) \to 1 \> (n \to \infty), \> P^{i,n-2k} _{k-1} = P^{i,n-2k} _{k-1}(2|a|^2-1) \> (i=0,1),$ and $I_A=1, I_G=-1.$
\par
Next we prepare the following asymptotic results for the Jacobi polynomial $P^{\alpha + an, \beta +bn} _n (x)$ derived by Chen and Ismail (1991): if $n \to \infty$ with $k/n=x \in (-(1-|a|)/2, (1+|a|)/2)$, then
\begin{eqnarray*}
&& P^{0,n-2k} _{k-1} \sim  
{ 2 |a|^{2k-n} \over \sqrt{\pi n \sqrt{- \Delta}} } \cos (An+B)  \\
&& P^{1,n-2k} _{k-1} \sim  
{2 |a|^{2k-n} \over \sqrt{\pi n \sqrt{- \Delta}}} \sqrt{{x \over (1-x)(1-|a|^2)}} \cos (An+B+ \theta) 
\end{eqnarray*}
where $\Delta = (1-|a|^2)(4x^2-4x+1-|a|^2)$, $A$ and $B$ are some constants (which are independent of $n$), and $\theta \in [0, \pi/2]$ is determined by $\cos \theta = \sqrt{(1-|a|^2)/4x(1-x)}$. 

\par
Combining the above observations with the Riemann-Lebesgue lemma (see p.462 in Durrett (1996)), we see that, if $n \to \infty$, then 

\begin{eqnarray*}
&& 
E(e^{i \xi {X_n ^{\varphi} \over n}}) 
\quad \to \quad  
\\
&&
{ 1-|a|^2 \over \pi} 
\int^{{1 \over 2}} _{{1 - |a| \over 2}} dx \> 
{ 1 \over x (1-x) \sqrt{(|a|^2-1) (4x^2-4x+1-|a|^2)}}   
\\
&&
\qquad \qquad \qquad
\times 
\left[
\cos ((1-2x) \xi)
-(1-2x) 
\left( 
|\alpha|^2 - |\beta|^2 +  
{\Theta_j \over |a|^2} 
\right) 
i \sin ((1-2x) \xi)
\right] 
\\
&&
=
\int^{|a|} _{-|a|} 
{ \sqrt{1 - |a|^2} \over \pi (1 - x^2) \sqrt{|a|^2 - x^2}} 
 \left\{ 1- \left( |\alpha|^2 - |\beta|^2 + { \Theta_j \over |a|^2 } \right) x \right\} 
e^{i \xi x}
\> 
dx
\\
&&
= \phi (\xi)
\end{eqnarray*}
Then $\phi (\xi)$ is continuous at $\xi =0$, so the continuity theorem (see p.99 in Durrett (1996)) implies that $X^{\varphi} _n/n$ converges in distribution to the $Z^{\varphi}$ with characteristic function $\phi$. Therefore Theorem 3 is obtained.

\par
Muraki (2002) introduced a notion of quasi-universal product for algebraic probability spaces and showed that there exist only five quasi-universal products, that is, tensor product (case 1), free product (case 2), Boolean product (case 3), monotone product (case 4), and anti-monotone product (case 5). Algebraic central limit theorems describe limit behaviors of rescaled sum of algebraic random variables $\sigma_1, \sigma_2, \ldots $ with mean 0 and variance 1,
\[
{ \sigma_1 + \sigma_2 + \cdots + \sigma_n \over \sqrt{n} }
\]
converges weakly, in the limit $n \to \infty$, to the Gaussian distribution $e^{-x^2/2}/\sqrt{2 \pi}$ (case 1), the semi-circle distribution $\chi_{[-2,2]}(x)\sqrt{4-x^2}/2\pi$ (case 2), the Bernoulli distribution $(\delta_{-1}+\delta_1)/2$ (case 3), and arcsine distribution $\chi_{(-\sqrt{2},\sqrt{2})}(x)/\pi \sqrt{2-x^2}$ (cases 4 and 5), where $\chi (A)=1$ if $x \in A$, $=0$ if $x \not\in A$ with $A \subset {\bf R}$ (see Muraki (2001), Hashimoto (2002), for examples). So our limit theorem (Theorem 3) would belong to another category which is different from ones given by Muraki.

\par
Now we compare our analytical results (given by Theorem 3) with the numerical ones for the A-type Haramard walk.

We see that Theorem 3 implies that if $ -\sqrt{2}/2 <a <b <\sqrt{2}/2$, then as $n \to \infty$, 
\begin{eqnarray*}
P(a \le X^{\varphi} _{A,n}/n \le b)  \to  \int^b _a {  1-  (|\alpha|^2 - |\beta|^2 + \Theta_A) x \over \pi (1-x^2) \sqrt{1-2x^2}} dx
\end{eqnarray*}
for any initial qubit state $\varphi = {}^t[\alpha, \beta]$. For the symmetric CRW $Y^o _n$ starting from the origin, the well-known central limit theorem implies that if $ - \infty <a <b < \infty $, then as $n \to \infty$, 
\begin{eqnarray*}
P(a \le Y^o _{n}/ \sqrt{n} \le b)  \to \int^b _a {e^{-x^2/2} \over \sqrt{2 \pi}} dx\end{eqnarray*}
This result is often called the de Moivre-Laplace theorem. When we take $\varphi = {}^t[1/\sqrt{2},i/\sqrt{2}]$ (symmetric case), then we have the following QRW version of the de Moivre-Laplace theorem: if $ -\sqrt{2}/2 <a <b <\sqrt{2}/2$, then as $n \to \infty$, 
\begin{eqnarray*}
P(a \le X^{\varphi} _{A,n}/n \le b) \to \int^b _a {1 \over \pi (1-x^2) \sqrt{1-2x^2}} dx
\end{eqnarray*}
So there is a difference between the QRW $X^{\varphi} _{A,n}$ and the CRW $Y^o _n$ even in a symmetric case for $\varphi = {}^t[1/\sqrt{2},i/\sqrt{2}]$. 

\par
Noting that $E(X^{\varphi} _{A,n})=0 \> (n \ge 0)$ for any $\varphi \in \Phi_{A,s}$, we have
\begin{eqnarray*}
sd(X_{A,n} ^{\varphi})/n \to  \sqrt{(2 - \sqrt{2})/2} = 0.54119 \ldots 
\end{eqnarray*}
where $sd(X)$ is the standard deviation of $X$. This rigorous result reveals that numerical simulation result 3/5 = 0.6 given by Travaglione and Milburn (2002) is not so accurate.
\par
As in a similar way, when we take $\varphi = {}^t[0,e^{i \theta}]$ where $\theta \in [0, 2\pi)$ (asymmetric case), we see that if $ -\sqrt{2}/2 <a <b <\sqrt{2}/2$, then as $n \to \infty$, 
\begin{eqnarray*}
P(a \le X^{\varphi} _{A,n}/n \le b)  \to  \int^b _a {1 \over \pi (1-x) \sqrt{1-2x^2}} dx
\end{eqnarray*}
So we have
\begin{eqnarray*}
E(X_{A,n} ^{\varphi})/n  \to  (2 - \sqrt{2})/2 = 0.29289 \ldots , \quad 
sd(X_{A,n} ^{\varphi})/n  \to \sqrt{(\sqrt{2} -1)/2} = 0.45508 \ldots 
\end{eqnarray*}
When $\varphi = {}^t[0,1]$ ($\theta =0$), Ambainis {\it et al.} (2001) gave the same result. In the paper, they took two approaches, that is, the Schr\"odinger approach and the path integral approach. However their result comes mainly from the Schr\"odinger approach by using a Fourier analysis. 
\par
In another asymmetric case $\varphi = {}^t[e^{i \theta},0]$ where $\theta \in [0, 2\pi)$, a similar argument implies that if $ -\sqrt{2}/2 <a <b <\sqrt{2}/2$, then as $n \to \infty$, 
\begin{eqnarray*}
P(a \le X^{\varphi} _{A,n}/n \le b)  \to  \int^b _a {1 \over \pi (1+x) \sqrt{1-2x^2}} dx
\end{eqnarray*}
Note that $f_A(-x ; {}^t[e^{i \theta},0])= f_A(x; {}^t[0, e^{i \theta}])$ for any $x \in (-\sqrt{2}/2, \sqrt{2}/2)$. Therefore concerning the $m$th moment of the limit distribution, we have the same result as in the previous case $\varphi = {}^t[0, e^{i \theta}]$. So the standard deviation of the limit distribution $Z_A ^{\varphi}$ is given by $\sqrt{(\sqrt{2} -1)/2} = 0.45508 \ldots.$ Simulation result 0.4544 $\pm$ 0.0012 in Mackay {\it et al.} (2002) (their case is $\theta =0$) is consistent with our rigorous result.

\section{Absorption Problem}
From now on we consider absorption problems for both A-type and G-type QRWs located on the sets $\{0,1, \ldots, N\}$ or $\{0,1, \ldots \}$. Results in this section for the A-type QRW appear in Konno, Namiki, Soshi and Sudbury (2003).

Before we move to a quantum case, first we describe the CRW on a finite set $\{0,1, \ldots, N\}$ with two absorption barriers at locations $0$ and $N$ (see Doyle and Snell (1984), Grimmett and Stirzaker (1992), for examples). The particle moves at each step either one unit to the left with probability $p$ or one unit to the right with probability $q=1-p$ until it hits one of the absorption barriers. The directions of different steps are independent of each other. The CRW starting from $k \in \{0,1, \ldots, N \}$ at time $n$ is denoted by $Y^k _n$ here. Let 
\[
T_m = \min \{ n \ge 0 : Y^k _n = m \}
\]
be the time of the first visit to $m \in \{0,1, \ldots, N \}$. Using the subscript $k$ to indicate $Y^k _0=k$, we let
\[
P^{(N)} _k = P_k (T_0 < T_N)
\]
be the probability that the particle hits $0$ starting from $k$ before it arrives at $N$. The absorption problem is also known as the Gambler's ruin problem. 
\par
Now we review some known results and conjectures on absorption problems related to this paper for A-type QRWs. 

In the case of $U=H$ (the Hadamard walk), when $N= \infty$, that is, the state space is $\{0,1, \ldots \}$ case, Ambainis {\it et al.} (2001) proved
\begin{eqnarray}
P^{(\infty)} _{A,1} ({}^t[0,1]) = P^{(\infty)} _{A,1} ({}^t[1,0]) = {2 \over \pi}
\end{eqnarray}
and Bach {\it et al.} (2002) showed 
\begin{eqnarray*} 
\lim_{k \to \infty} P^{(\infty)} _{A,k} (\varphi) 
=  \left({1 \over 2} \right) |\alpha|^2 
+ \left( {2 \over \pi} - {1 \over 2} \right)|\beta|^2
+ 2 \left( {1 \over \pi} - {1 \over 2} \right) \Re(\overline{\alpha}\beta)
\end{eqnarray*}
for any initial qubit state $\varphi ={}^t [\alpha, \beta] \in \Phi.$ Furthermore, in the case of $U=H(\rho)$, Bach {\it et al.} (2002) gave
\begin{eqnarray*} 
&& \lim_{k \to \infty} P^{(\infty)} _{A,k} ({}^t[0,1]) 
= {\rho \over 1- \rho} \left( {\cos^{-1} (1- 2\rho) \over \pi} -1 \right) 
+ { 2 \over \pi \sqrt{1/\rho -1}} \\
&& \lim_{k \to \infty} P^{(\infty)} _{A,k} ({}^t[1,0]) 
= {\cos^{-1} (1- 2\rho) \over \pi} 
\end{eqnarray*}  
The second result was conjectured by Yamasaki, Kobayashi and Imai (2002).

When $N$ is finite, the following conjecture by Ambainis {\it et al.} (2001) is still open for the $U=H$ case:
\begin{eqnarray*} 
P^{(N+1)} _{A,1} ({}^t[0,1]) ={ 2 P^{(N)} _{A,1} ({}^t[0,1]) +1 \over 2 P^{(N)} _{A,1} ({}^t[0,1]) +2} \quad (N \ge 1), \qquad P^{(1)} _{A,1} ({}^t[0,1]) =0. 
\end{eqnarray*} 
Solving the above recurrence gives
\begin{eqnarray} 
P^{(N)} _{A,1} ({}^t[0,1]) = {1 \over \sqrt{2}} \times {(3+2 \sqrt{2})^{N-1} -1 \over (3+2 \sqrt{2})^{N-1} +1} \quad (N \ge 1)
\end{eqnarray}
However in contrast with $P^{(\infty)} _{A,1} ({}^t[0,1]) =2/\pi$, Ambainis {\it et al.} (2001) proved
\[ 
\lim_{N \to \infty} P^{(N)} _{A,1} ({}^t[0,1])=1/\sqrt{2}
\]

Let $T_0$ be the first hitting time to 0. We consider the conditional $m$th moment of $T_0$ starting from $k=1$ given the event $\{T_0 < \infty \}$, that is, $E_{j,1} ^{(\infty)}((T_0)^m|T_0 < \infty)= E_{j,1} ^{(\infty)} ((T_0)^m ; T_0 < \infty)/P_{j,1} ^{(\infty)} ( T_0 < \infty)$ for the $j$-type QRW. 

From now on we begin with the classical case. In this case, to obtain $P^{(N)} _k$, we use the following difference equation:
\begin{eqnarray}
P^{(N)} _k = p P^{(N)} _{k-1} + q P^{(N)} _{k+1} \qquad (1 \le k \le N-1)
\end{eqnarray}
with boundary conditions:
\begin{eqnarray}
P^{(N)} _0=1, \quad  P^{(N)} _N=0
\end{eqnarray}
To consider a similar equation even in the quantum case and to use the PQRS method are our basic strategy.

From now on we focus on $1 \le k \le N-1$ case. So we consider only $n \ge 1$ case. Noting that $\{ P_j,Q_j,R_j,S_j \}$ for each $j=A,G$ is a basis of $M_2 ({\bf C})$, $\Xi^{(N)} _{j,k} (n)$ can be written as
\begin{eqnarray*} 
\Xi^{(N)} _{j,k} (n) = p^{(N)} _{j,k} (n) P_j + q^{(N)} _{j,k} (n) Q_j + r^{(N)} _{j,k} (n) R_j + s^{(N)} _{j,k} (n) S_j 
\end{eqnarray*} 
From the definition of $\Xi^{(N)} _{j,k} (n)$, it is easily shown that there exist only two types of paths, that is, $P_j \ldots P_j$ and $P_j \ldots Q_j$, since we consider only a hitting time to $0$ before it arrives at $N$. Therefore we see that $q^{(N)} _{j,k} (n)=s^{(N)} _{j,k} (n)=0 \> (n \ge 1)$.

We assume $N \ge 3$. Noting that the definition of $\Xi^{(N)} _{j,k} (n)$, we have 
\begin{eqnarray*} 
\Xi^{(N)} _{j,k} (n) = \Xi^{(N)} _{j,k-1} (n-1)P_j + \Xi^{(N)} _{j,k+1} (n-1)Q_j  \qquad (1 \le k \le N-1)
\end{eqnarray*} 
The above equation is a QRW version of the difference equation, i.e., Eq. (4.7) for the CRW. As an example, see Eq. (2.4). Then we have
\begin{eqnarray*} 
&& p^{(N)} _{j,k} (n) = a p^{(N)} _{j,k-1} (n-1)+ c r^{(N)} _{j,k-1} (n-1)\\ 
&& r^{(N)} _{j,k} (n) = b p^{(N)} _{j,k+1} (n-1)+ d r^{(N)} _{j,k+1} (n-1)
\end{eqnarray*} 
Note that the above equations do not depend on types of QRWs. Next we consider boundary conditions related to Eq. (4.8) in the classical case. When $k=N$, 
\[
P^{(N)} _{j,N} (0;\varphi) = | \Xi^{(N)} _{j,N} (0) \varphi |^2 =0
\]
for any $\varphi \in \Phi$. So we take $\Xi^{(N)} _{j,N} (0) = O_2$. In this case, Eq. (2.2) gives   
\[
p^{(N)} _{j,N} (0) =  r^{(N)} _{j,N} (0) = 0
\]
If $k=0$, then
\[
P^{(N)} _{j,0} (0;\varphi) = | \Xi^{(N)} _{j,0} (0) \varphi |^2 =1
\]
for any $\varphi \in \Phi$. So we choose $\Xi^{(N)} _{j,N} (0) = I_2$. From Eq. (2.4), we have
\[
p^{(N)} _{j,0} (0) = \overline{a}, \>\> r^{(N)} _{j,0} (0) = \overline{c}
\]
It is noted that the above boundary conditions also do not depend on types of QRWs. Therefore, from now on we will omit subscript $j$ of $p^{(N)} _{j,k} (n)$ and $r^{(N)} _{j,k} (n).$ Let
\begin{eqnarray*} 
v^{(N)} _{k} (n) = 
\left[
\begin{array}{cc}
p^{(N)} _{k} (n) \\
r^{(N)} _{k} (n)   
\end{array}
\right] 
\end{eqnarray*}
Then we see that for $n \ge 1$ and $1 \le k \le N-1$, 
\begin{eqnarray} 
&& v^{(N)} _{k} (n) = 
\left[
\begin{array}{cc}
a & c \\
0 & 0 
\end{array}
\right] 
v^{(N)} _{k-1} (n-1)
+
\left[
\begin{array}{cc}
0 & 0 \\
b & d 
\end{array}
\right] 
v^{(N)} _{k+1} (n-1)
\end{eqnarray}
and for $1 \le k \le N$, 
\begin{eqnarray*}
&& v^{(N)} _{0} (0) 
= 
\left[
\begin{array}{cc}
\overline{a}  \\
\overline{c}
\end{array}
\right] 
, \quad
v^{(N)} _{k} (0) 
= 
\left[
\begin{array}{cc}
0  \\
0
\end{array}
\right] 
\end{eqnarray*}
Moreover,
\begin{eqnarray*}
v^{(N)} _{0} (n) 
=
v^{(N)} _{N} (n) 
= 
\left[
\begin{array}{cc}
0  \\
0
\end{array}
\right] 
\quad
(n \ge 1)
\end{eqnarray*}
So the definition of $P^{(N)} _{j,k} (\varphi)$ gives

\begin{lem}
\label{lem:lem1}
\begin{eqnarray*} 
&& P^{(N)} _{j,k} (\varphi) = \sum_{n=1} ^{\infty} P^{(N)} _{j,k} 
(n; \varphi) \\
&& P^{(N)} _{j,k} (n; \varphi) 
= C_{j,1} (n) |\alpha|^2 + C_{j,2} (n) |\beta|^2 + 
2 \Re(C_{j,3} (n) \overline{\alpha} \beta)
\end{eqnarray*} 
where $\Re (z)$ is the real part of $z \in {\bf C}$, $\varphi = {}^t[\alpha, \beta] \in \Phi$ and 
\begin{eqnarray*}
&& C_{A,1} (n) = |a p^{(N)} _k (n) + c r^{(N)} _k (n)|^2, \quad
C_{A,2} (n) = |b p^{(N)} _k (n) + d r^{(N)} _k (n)|^2  \\
&& C_{A,3} (n) = 
\overline{(a p^{(N)} _k (n) + c r^{(N)} _k (n))} (b p^{(N)} _k (n) + d r^{(N)} _k (n)) \\
&& C_{G,1} (n) = |p^{(N)} _k (n)|^2, \quad
C_{G,2} (n) = |r^{(N)} _k (n)|^2, \quad 
C_{G,3} (n) = 
\overline{p^{(N)} _k (n)} r^{(N)} _k (n)
\end{eqnarray*}
\end{lem}

To solve $P^{(N)} _{j,k} (\varphi)$, we introduce generating functions of $p^{(N)} _k (n)$ and $r^{(N)} _k (n)$ as follows:
\begin{eqnarray*}
\widetilde{p}^{(N)} _k (z) = \sum_{n=1} ^{\infty} p^{(N)} _k (n) z^n,
\qquad  \widetilde{r}^{(N)} _k (z) = \sum_{n=1} ^{\infty} r^{(N)} _k (n) z^n
\end{eqnarray*}
By Eq. (4.9), we have
\begin{eqnarray*} 
&& \widetilde{p}^{(N)} _k (z) = a z \widetilde{p}^{(N)} _{k-1} (z) + cz \widetilde{r}^{(N)} _{k-1} (z) \\
&& \widetilde{r}^{(N)} _k (z) = b z \widetilde{p}^{(N)} _{k+1} (z) + dz \widetilde{r}^{(N)} _{k+1} (z) 
\end{eqnarray*} 
Solving these, we see that both $\widetilde{p}^{(N)} _k (z)$ and $\widetilde{r}^{(N)} _k (z)$ satisfy the same recurrence:   
\begin{eqnarray*} 
&& d \widetilde{p}^{(N)} _{k+2} (z) - \left( \triangle z+{1 \over z} \right) \widetilde{p}^{(N)} _{k+1} (z) + a \widetilde{p}^{(N)} _{k} (z) = 0\\
&& d \widetilde{r}^{(N)} _{k+2} (z) - \left( \triangle z+{1 \over z} \right) \widetilde{r}^{(N)} _{k+1} (z) + a \widetilde{r}^{(N)} _{k} (z) = 0
\end{eqnarray*} 
From the characteristic equations with respect to the above recurrences, we have the same root: if $a \not= 0$, then
\begin{eqnarray*} 
\lambda_{\pm} = {\triangle z^2 + 1 \mp \sqrt{\triangle^2 z^4 + 2 \triangle ( 1- 2 |a|^2)z^2 + 1} \over 2 \triangle \overline{a} z} 
\end{eqnarray*} 
where $\triangle = \det U = ad - bc$.

From now on we consider mainly $U=H$ (the Hadamard walk) case with $N=\infty$. Remark that the definition of $\Xi_{j,1} ^{(\infty)} (n)$ gives $p_1 ^{(\infty)}(n)=0 \> (n \ge 2)$ and $p_1 ^{(\infty)}(1)=1$. So we have $\widetilde{p}^{(\infty)} _1 (z) = z$. Moreover noting $\lim_{k \to \infty} \widetilde{p}^{(\infty)}  _k(z) < \infty$, the following explicit form is obtained:
\begin{eqnarray*}
\widetilde{p}^{(\infty)} _k (z) = z \lambda_ +^{k-1}, \qquad 
\widetilde{r}^{(\infty)} _k (z) = \frac{-1+\sqrt{z^4+1}}{z} \lambda_+^{k-1}
\end{eqnarray*}
where $\lambda_\pm = (z^2-1\pm\sqrt{z^4+1})/{\sqrt{2}z}.$ Therefore for $k=1$, we have $\widetilde{r}^{(\infty)} _1 (z)= (-1+\sqrt{z^4+1})/z$. By using these, we obtain

\begin{pro}
\label{pro:pro1}
For each $j$-type QRW, we have
\begin{eqnarray}  
&&
P^{(\infty)} _{A,1} (\varphi) = {2 \over \pi} + 
2 \left(1 - {2 \over \pi} \right) \Re(\overline{\alpha} \beta), \\
&&
P^{(\infty)} _{G,1} (\varphi) = |\alpha|^2 + 
\left({4 \over \pi} - 1 \right) |\beta|^2
\end{eqnarray} 
for any initial qubit state $\varphi = {}^t [\alpha, \beta] \in \Phi$. 
\end{pro}
Eqs. (4.10) and (4.11) are a generalization of Eq. (4.5) given by Ambainis {\it et al.} (2001). Then we get $(4-\pi)/\pi \le P_{j,1} ^{(\infty)}(\varphi) \le 1$ for each $j$-type QRW. Moreover we obtain a result on the conditional $m$th moment of the first hitting time to 0 starting from $k=1$ given an event $\{T_0 < \infty \}$, that is, $E_{j,1} ^{(\infty)}((T_0)^m|T_0 < \infty)=E_{j,1} ^{(\infty)} ((T_0)^m ; T_0 < \infty)/P_{j,1} ^{(\infty)} ( T_0 < \infty)$ in a similar way. 

\begin{pro}
\label{pro:pro2} 
For each $j$-type QRW, we have
\begin{eqnarray*}
&& E_{j,1} ^{(\infty)} (T_0|T_0 < \infty) = {1 \over P_{j,1} ^{(\infty)}(\varphi)}
\\
&&
E_{j,1} ^{(\infty)} ((T_0)^m |T_0 < \infty) = \infty \qquad (m \ge 2)
\end{eqnarray*}
\end{pro}

Next we consider a finite $N$ case. By using boundary conditions: $\widetilde{p}^{(N)}_1(z)=z$ and $\widetilde{r}^{(N)} _{N-1} (z)=0$, we see that $\widetilde{p}^{(N)} _k (z)$ and $\widetilde{r}^{(N)} _k (z)$ satisfy 
\begin{eqnarray}
&& \widetilde{p}^{(N)}_k(z) = \left( {z \over 2} +E_z \right) \lambda_+^{k-1}
+ \left( {z \over 2} -E_z \right) \lambda_-^{k-1} \\
&& \widetilde{r}^{(N)}_k(z) = C_z (\lambda_+^{k-N+1}-\lambda_-^{k-N+1})
\end{eqnarray}
where 
\begin{eqnarray} 
&& C_z = {z^2 \over \sqrt{2}} 
(-1)^{N-2}(\lambda_+^{N-3} - \lambda_-^{N-3}) \\
&& \times \left\{ 
(\lambda_+^{N-2} -\lambda_-^{N-2})^2
-{z \over \sqrt{2}} (\lambda_+^{N-2} -\lambda_-^{N-2})
(\lambda_+^{N-3} -\lambda_-^{N-3})
-(-1)^{N-3}(\lambda_+ - \lambda_-)^2 \right\}^{-1} \nonumber \\
&& E_z = - {z \over 2(\lambda_+^{N-2} -\lambda_-^{N-2})}
\bigg[ 2 (-1)^{N-3}(\lambda_+ - \lambda_-) 
(\lambda_+^{N-3} -  \lambda_-^{N-3}) \\
&& \times \left\{ 
(\lambda_+^{N-2} -\lambda_-^{N-2})^2
-{z \over \sqrt{2}} (\lambda_+^{N-2} -\lambda_-^{N-2})
(\lambda_+^{N-3} -\lambda_-^{N-3})
-(-1)^{N-3} (\lambda_+ - \lambda_-)^2 \right\}^{-1} \nonumber \\
&& \qquad \qquad \qquad \qquad \qquad \qquad \qquad \qquad \qquad 
+ \> (\lambda_+^{N-2} + \lambda_-^{N-2}) \bigg] \nonumber 
\end{eqnarray} 
So we obtain 
\begin{thm}
\label{thm:thm2}
For each $j$-type QRW, we have
\begin{eqnarray*} 
P^{(N)} _{j,k} (\varphi) 
= C_{j,1} |\alpha|^2 + C_{j,2} |\beta|^2 + 2 \Re(C_{j,3} \overline{\alpha} \beta)
\end{eqnarray*} 
where $\varphi = {}^t[\alpha, \beta] \in \Phi$ and 
\begin{eqnarray*}
&& C_{A,1} = {1 \over 2 \pi} \int_0 ^{2\pi} |a \widetilde{p}^{(N)} _k (e^{i \theta}) + c \widetilde{r}^{(N)} _k (e^{i \theta})|^2 d \theta \\
&& C_{A,2} = {1 \over 2 \pi} \int_0 ^{2\pi} |b \widetilde{p}^{(N)} _k (e^{i \theta}) + d \widetilde{r}^{(N)} _k (e^{i \theta})|^2 d \theta \\
&& C_{A,3} =  {1 \over 2 \pi} \int_0 ^{2\pi}
\overline{(a \widetilde{p}^{(N)} _k (e^{i \theta}) + c \widetilde{r}^{(N)} _k (e^{i \theta}))} (b \widetilde{p}^{(N)} _k (e^{i \theta}) + d \widetilde{r}^{(N)} _k (e^{i \theta}) ) d \theta \\
&& C_{G,1} = {1 \over 2 \pi} \int_0 ^{2\pi} |\widetilde{p}^{(N)} _k (e^{i \theta})|^2 d \theta, 
\quad 
C_{G,2} = {1 \over 2 \pi} \int_0 ^{2\pi} |\widetilde{r}^{(N)} _k (e^{i \theta})|^2 d \theta \\
&& C_{G,3} =  {1 \over 2 \pi} \int_0 ^{2\pi}
\overline{\widetilde{p}^{(N)} _k (e^{i \theta})} \widetilde{r}^{(N)} _k (e^{i \theta}) d \theta 
\end{eqnarray*}
with $a=b=c=-d=1/\sqrt{2}$, here $\widetilde{p}^{(N)} _k (z)$ and $\widetilde{r}^{(N)} _k (z)$ satisfy Eqs. (4.12) and (4.13), and 
$C_z$ and $E_z$ satisfy Eqs. (4.14) and (4.15).
\end{thm}

Here we consider $U=H$ (the Hadamard walk), $\varphi = {}^t [\alpha, \beta]$ and $k=1$. From Theorem 8, noting that $\widetilde{p}^{(N)} _1 (z)= z$ for any $N \ge 2$, we have 
\begin{cor}
\label{cor:cor3} For each $j$-type QRW, if $N \ge 2$, then 
\begin{eqnarray*}
&&
P^{(N)} _{A,1} (\varphi) 
={1 \over 2} \left( 1 + {1 \over 2 \pi} \int_0 ^{2\pi} |\widetilde{r}^{(N)} _1 (e^{i \theta})|^2 d \theta \right) 
( 1 + 2 \Re(\overline{\alpha} \beta))
\\
&&
P^{(N)} _{G,1} (\varphi) 
=
|\alpha|^2
+ \left( 
{1 \over 2 \pi} \int_0 ^{2\pi} |\widetilde{r}^{(N)} _1 (e^{i \theta})|^2 d \theta 
\right) |\beta|^2
\end{eqnarray*}
where $\widetilde{r}^{(2)} _1 (z) = 0, \> \widetilde{r}^{(3)} _1 (z) = z^3/(2 - z^2)$ and for $N \ge 4$
\begin{eqnarray*}
\widetilde{r}^{(N)} _1 (z)= - 
{ z^2 J_{N-3} (z) J_{N-4} (z) \over \sqrt{2} (J_{N-3} (z))^2 -z J_{N-3} (z) J_{N-4} (z) - \sqrt{2} (-1)^{N-3}} 
\end{eqnarray*}
with 
\[
J_n (z) = \sum_{k=0} ^n \lambda_+ ^k \lambda_- ^{n-k}, \quad \lambda_+ + \lambda_- = \sqrt{2} \left( z - {1 \over z} \right), \quad \lambda_+ \lambda_- = -1
\]
\end{cor}
In particular, when $\varphi = {}^t [0,1] = |R \rangle, \> k=1$ and $N=2, \ldots ,6$ cases, the above corollary for the A-type QRW implies that the values $P^{(N)}_{A,1} ({}^t [0,1] ) \> (N=2, \ldots ,6)$ satisfy the conjecture given by Eq. (4.6).

\par
\
\par\noindent
{\bf Acknowledgments.}  This work is partially financed by the Grant-in-Aid for Scientific Research (B) (No.12440024) of Japan Society of the Promotion of Science. I would like to thank Takao Namiki, Takahiro Soshi, Hideki Tanemura, and Makoto Katori for useful discussions.

\par
\
\par\noindent

\begin{small}

\bibliographystyle{plain}

\par
\vskip 1.0cm

Department of Applied Mathematics

Faculty of Engineering

Yokohama National University

Hodogaya-ku, Yokohama 240-8501, Japan

norio@mathlab.sci.ynu.ac.jp

\vskip 1cm

\end{small}

\end{document}